\documentclass{amsproc}
\usepackage[applemac]{inputenc}
\usepackage{graphicx}
\usepackage[frenchb,english]{babel}

\theoremstyle{definition}

\theoremstyle{remark}

\numberwithin{equation}{section}

\begin{document}
\title{Diffusive chaos in navigation satellites orbits}
%    Remove any unused author tags.

%    author one information
\author{J.\,Daquin$^{*,\dag,\ddag}$, A.J.\,Rosengren$^{\sharp,\natural}$, K.\,Tsiganis$^{\natural}$}

\address{$^\dag$Space Research Centre, School of Sciences\\
The RMIT University, Melbourne $3001$, Australia.
\newline
$^*$E-mail: jerome.daquin@rmit.edu.au.
}

\address{$^\ddag$IMCCE, $77$ Avenue Denfert-Rochereau\\
Paris, $75014$,  France.
}

\address{$^\sharp$IFAC-CNR, Via Madonna del Piano $10$\\
Sesto Fiorentino, $50019$, Italy.
}

\address{$^\natural$Department of Physics, Aristotle University of Thessaloniki\\
Thessaloniki, $54124$, Greece.
}

%\thanks{Special thanks are due to}
%    author two information
%\author{}
%\address{}
%\curraddr{}
%\email{}
%\thanks{}
\begin{abstract}

The navigation satellite constellations in medium-Earth orbit exist in a background of third-body secular resonances stemming from the perturbing gravitational effects of the Moon and the Sun. The resulting chaotic motions, emanating from the overlapping of neighboring resonant harmonics, induce especially strong perturbations on the orbital eccentricity, which can be transported to large values, thereby increasing the collision risk to the constellations and possibly leading to a proliferation of space debris. We show here that this transport is of a diffusive nature and we present representative diffusion maps that are useful in obtaining a global comprehension of the dynamical structure of the navigation satellite orbits. 

\end{abstract}
%\subjclass[2010]{Primary }
\keywords{Orbital resonances; Chaotic diffusion; Secular dynamics; Medium-Earth orbits; Navigation satellites.}

\maketitle

%-------------------------------------
\section{Introduction}\label{sec:sec1}
%-------------------------------------

The past several years have seen a renewed interest in the dynamics of medium-Earth orbits (MEOs), the region of the navigation satellites\footnote{\textit{i.e}, semi-major axes between $3$ and $5$ Earth radii.},  since the community has realized the inherent dangers imposed by space debris, meanwhile stimulating a deeper dynamical understanding of this multifrequency and variously perturbed environment. The effects of the Moon and the Sun on Earth-orbiting satellites, often negligible on short timescales, may have profound consequences on the motion over longer periods; this accumulating effect is a phenomenon known as resonance. The inclined, nearly circular orbits of the navigation satellites are not excluded from this situation.

\begin{figure}
\begin{center}
\includegraphics[scale=0.4]{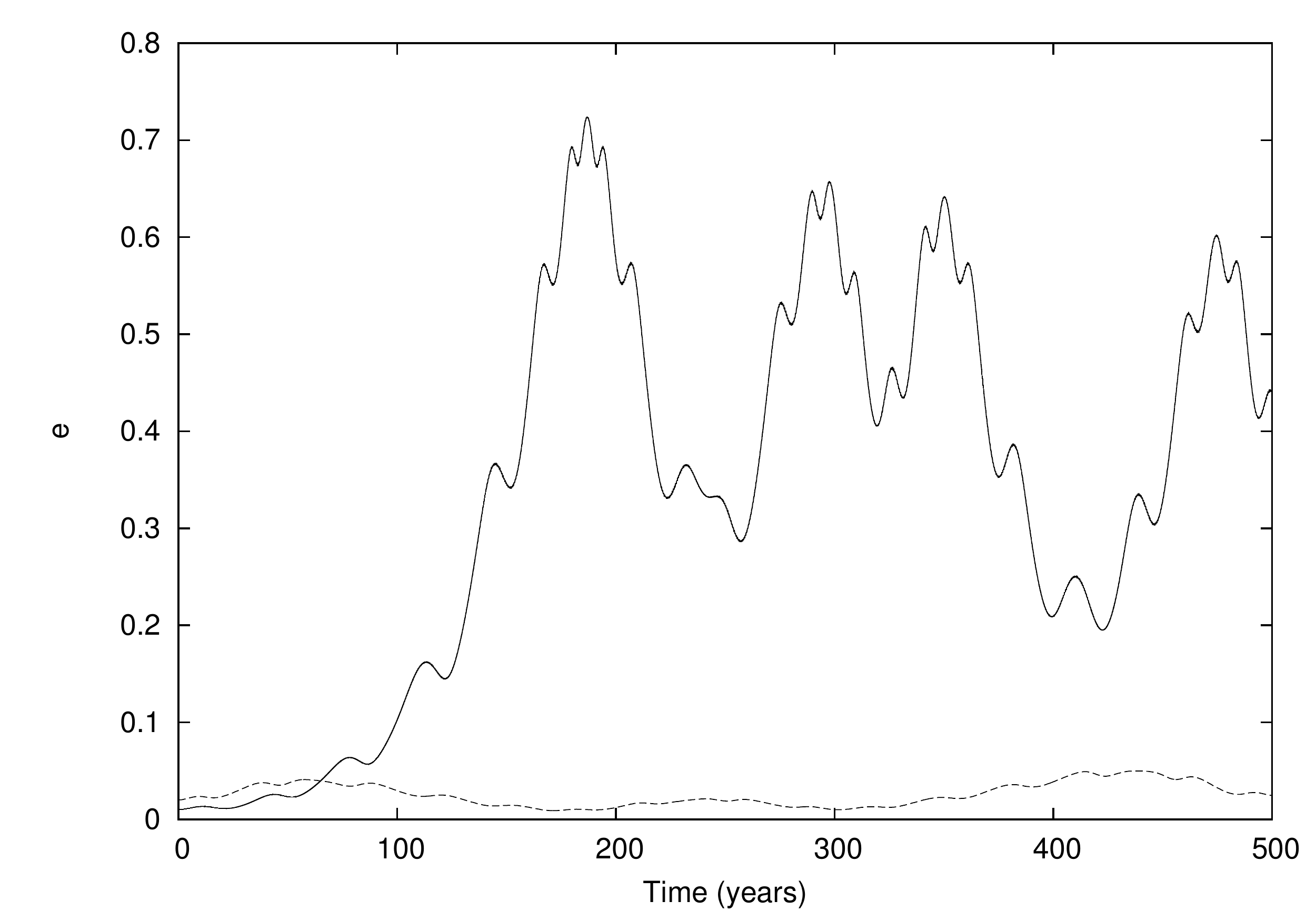}
\end{center}
\caption{\label{fig:e-versus-time}Typical eccentricity history for orbits in the MEO region: the orbit with a large variation has been declared as chaotic by the FLI analysis while the orbit with modest excursions is a regular non-resonant orbit.
Note that the eccentricity can be transported to large values in the chaotic case.}
\end{figure}

Several, but mainly numerical, works \cite{tEl97,fDe11,aRo08} have quickly pointed out the key role played by the lunar and solar third-body resonances, especially on the orbital eccentricity. This instability manifests itself as an apparent chaotic growth of the eccentricity on decadal timescales, as illustrated by Fig.\,\ref{fig:e-versus-time}. Here, the orbits have been numerically integrated using an in-house, high-precision orbit propagation code, based on classical averaging formulations of the equations of motion---a well-known and efficient technique for treating long-term evolutions in celestial mechanics\footnote{More precisely, the state vector $x \in \mathbb{R}^{6}$  of the satellite has been decomposed in terms of mean elements, $\bar{x} \in \mathbb{R}^{6}$, plus a small remainder as $x_{i} = \bar{x}_{i} + \xi_{i}(x,t), \ i=1,\cdots,6$, where the vector $\bar{x}$ obeys the standard form of perturbed systems, separated in slow-fast variables: 
\begin{align}
\left\{
	\begin{aligned} 
 \bar{x}_{i} &= \epsilon f_{i}(\bar{x}_{1},\cdots,\bar{x}_{5},t), \ i=1,\cdots,5 \notag \\
 \bar{x}_{6}&= n +  \epsilon f_{6}(\bar{x}_{1},\cdots,\bar{x}_{5},t).
	\end{aligned}
	\right.	
\end{align}
The quantity $n$ denotes here the  (mean) mean-motion. 
Short-periodic variations  are only present in the $\xi_{i}, \ i=1,\cdots,6,$ terms.
Due to the fact that $(\bar{x}_{1},\cdots,\bar{x}_{5})$ are slow variables, large step size can be used to propagate numerically the dynamics, useful for long-term ephemeris calculations and predictions.}.
Using a first-oder variational stability indicator, the fast Lyapunov indicator (FLI) \cite{cFr97,cFr00}, these orbits have been declared \textit{a posteriori} as chaotic and regular non-resonant.

On the analytical point of view, it is regrettable to note that no real effort was hitherto made in the literature  to guide the problem towards a global  comprehension of the observed instabilities, capturing in the same time the (supposed) dynamical richness of the inclination-eccentricity ($i~-~e$) phase-space. The complexity of this perturbed dynamical environment, however, is now becoming more clear \cite{aRo15,jDa15}.

The present work summarize our latest results towards understanding the chaotic structures of the phase-space near the lunisolar resonances. In particular, we show that the transport properties of the eccentricity in the phase-space, due to chaos, are of a diffusive nature, and we present some results on the numerical estimates of the diffusion coefficient relevant to navigation satellite parameters, especially for the European Galileo constellation.  

%---------------------------------
% e, FLI, D versus time

%-------------------------------------
\section{The dynamics of MEOs}
%-------------------------------------
We review the main features and the recent results that we have obtained for the dynamical description of the MEO region. This sections emphasizes ideas, but, for the sake of brevity, not the rigor of all the details involved.  
%-------------------------------------
\subsection{Overlap of the lunisolar secular resonances \textit{\`a la Chirikov}}
%-------------------------------------
The Hamiltonian system, written in canonical action-angles variables, is a small perturbation of an integrable system, 
\begin{align}
\mathcal{H} = \mathcal{H}_{0}(I) + \epsilon  \mathcal{H}_{1}(I,\Phi), \ (I,\Phi) \in \mathbb{R}^{9} \times \mathbb{T}^{9},  \epsilon \ll 1;
\end{align}
namely of the Kepler two-body problem. Considering a mathematically simple, but physically relevant dynamical model, the Hamiltonian governing the dynamics is composed of the Keplerian part of the geopotential, the oblateness  effect of the Earth, and the gravitational perturbations of third bodies, \textit{i.e.}, the Moon and the Sun: 
\begin{align}\label{Hbm}
	\mathcal{H}=\mathcal{H}_{\rm{kep.}} + \mathcal{H}_{J_{2}} + \mathcal{H}_{\rm{M}} + \mathcal{H}_{\rm{S}}.
\end{align}
Explicit and detailed formulas for these terms can be found in several works \cite{tEl97,aRo15,jDa15}.
The secular Hamiltonian, a $2.5$ degree-of-freedom (DOF) system\footnote{i.e., $2$-DOF and non-autonomous.}, useful for describing the long-term dynamics, has been derived and reduced, treating the resonances in isolation, from Eq.\,\eqref{Hbm} to the \textit{first fundamental model of resonances} \cite{sBr03} (a pendulum) near each resonance by constructing suitable (canonical) resonant variables \cite{jDa15}. 
These lunisolar secular resonances involve a linear combination of $2$ angles of the satellite, the argument of perigee $\omega$ and the ascending node $\Omega$, combined with the ascending node of the Moon $\Omega_{\rm M}$, which satisfy the resonant condition
\begin{align}\label{resonantcondition}
	\dot{\sigma}_{\bold{n}} \equiv n_{1} \dot{\omega} + n_{2} \dot{\Omega} + n_{3} \dot{\Omega}_{\rm M} \sim 0, \ \bold{n}=(n_{1},n_{2},n_{3}) \in \mathbb{Z}^{3}_{\star}.
\end{align}
The resonance centers $\mathcal{C}_{\bold{n}}$ are located in the action phase-space by the actions satisfying the equality $\dot{\sigma}_{\bold{n}}=0$. Going back to the eccentricity--inclination variables, which are physically and geometrically more interpretable (especially to space engineers), it can be shown that condition \eqref{resonantcondition} is equivalent to the relation $f_{\bold{n},a_{\star}}(e,i)=0$, with
\begin{align}
f_{\bold{n},a_{\star}}(e,i): [0,1] \times [0,2\pi] \rightarrow \mathbb{R};
\end{align}
a function parametrized by the the initial semi-major axis $a_{\star}$, a free parameter of the problem\footnote{In the secular version of the Hamiltonian, the canonical angle `associated' to the semi-major axis is a cyclic variable, so that the semi-major axis is a first integral \cite{jDa15}.}. The resonance centers  $\mathcal{C}_{\bold{n}}$,
\begin{align}\label{eq:centers}
	\mathcal{C}_{\bold{n}} \equiv \big\{ 
	(e,i) \in [0,1] \times [0,2\pi] \ \big\vert \ \big(\dot{\sigma}_{\bold{n}} = 0\big) \Leftrightarrow \big(f_{\bold{n},a_{\star}}(e,i)=0\big)
	\big\},
\end{align}
form a dense network of curves in the $i$--$e$ phase-space. When $a_{\star}$ is receding from $3$ to $5$ Earth radii, sweeping the navigation constellation regime, the resonance curves began to intersect, indicating locations where several critical arguments $\sigma_{\bold{n}}$ have 
vanishing frequencies simultaneously. 

\begin{figure}
\centering
\includegraphics[scale=1]{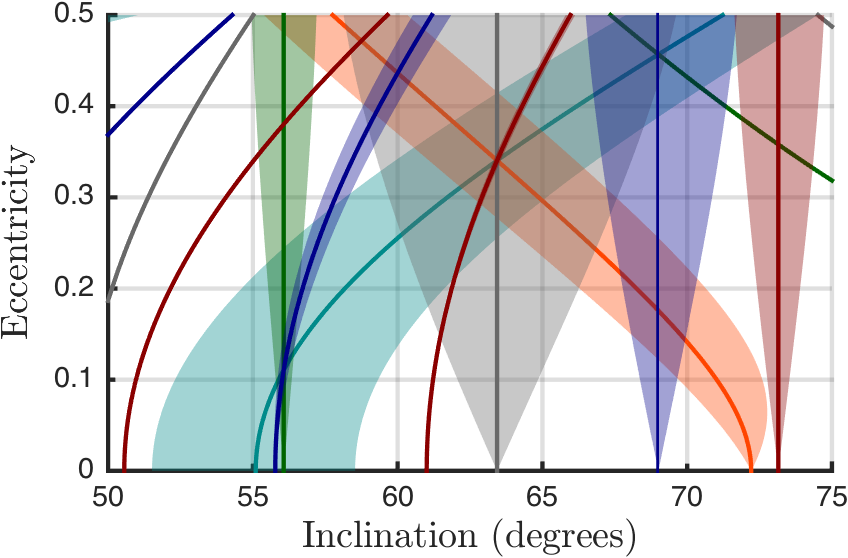}
\caption{\label{fig:chirikov} Lunisolar resonance centers $\mathcal{C}_{\bold{n}}$ (solid lines) and widths $\mathcal{W}^{\pm}_{\bold{n}}$ (transparent shapes) for $a_{\star}=29,600$km, \textit{i.e.}, Galileo's nominal semi-major axis. This plot shows the overlap between the first resonant harmonics $(\vert n_{i} \vert \le 2, i=1,\cdots,3)$. Galileo satellites are located near $i=56^{\circ}$.}
\end{figure}

Treating each resonances in isolation, and using the fundamental reduction to the pendulum, the amplitudes $\Delta_{\bold{n}}$ of each resonance associated to the critical argument $\sigma_{\bold{n}}$ have been estimated. The `maximal excursion' curves in the $i~-~e$ phase-space, delimiting the resonant domains, are then defined as 
 \begin{align}
	\mathcal{W}_{\bold{n}}^{\pm} \equiv \big\{ 
	(e,i) \in [0,1] \times [0,2\pi] \ \big\vert \ \dot{\sigma}_{\bold{n}} = \pm \Delta_{\bold{n}}
	\big\}.
\end{align}
We found a transition from a `stability regime', where resonances are thin and well separated at $a_{\star}=19,000$ km ($\sim 3$ Earth radii), to a `Chirikov one' \cite{bCh79}, where resonances overlap significantly at $a_{\star}=29,600$ km ($\sim 4.6$ Earth radii), the initial semi-major axis of the European navigation constellation, Galileo, as illustrated in Fig.\,\ref{fig:chirikov}. This important structural and dynamical fact has been obscured for nearly 2 decades, despite the pioneering breakthroughs of T. Ely \cite{tEl97,tEl02}. The analytical Chirikov resonance-overlap criterion that we applied was tested with respect to a detailed numerical FLI analysis of the phase-space, producing a stability atlas, a collection of FLI maps. The FLI analysis has confirmed the existence of the complex stochastic regime, whose effects on the dynamics is of primary importance \cite{jDa15}.

%-------------------------------------
\subsection{Transport in action space}
%-------------------------------------
Since the famous example of the asteroid Helga in Milani and Nobili's work \cite{aMi92}, physical orbits in the Solar System can be much more stable than their characteristic Lyapunov times would suggest, a concept referred to as \textit{stable chaos}. Thus, understanding the physical manifestation (the signature) of chaos on the system is preeminent. 
Rosengren et al. have recently demonstrated that the transport phenomenon acting in phase-space is intimately related to the resonant skeleton described by the centers $\mathcal{C}_{\bold{n}}$ \cite{aRo15}, confirming Ely's original results \cite{tEl02}, but on a much shorter timescale. They showed \textit{via} a discretization of the dynamics (stroboscopic approaches) that the transport in the phase-space is mediated by the web-like structures of the secular resonance centers $\mathcal{C}_{\bold{n}}$, allowing nearly circular orbits to become highly elliptic (as already illustrated by Fig.\,\ref{fig:e-versus-time}). 
This idea was further enlivened, taking advantages of the geometry and topology of the chaotic structures revealed by our FLI analysis. In fact, we showed that the long-term evolution of chaotic orbits superimposed on the background dynamical structures obtained \textit{via} the FLIs tends to evolve in the chaotic sea, exploring consequently a large phase-space volume. This is contrarily to stable orbits whose excursion in eccentricity and inclination are much more modest, being confined by KAM curves. Thus, in addition to quantifying the local hyperbolicity, the FLI maps also reveal how the transport is mediated in the phase-space, revealing the preferential routes of transport \cite{jDa15}.

%-------------------------------------
\section{Diffusive chaos}
%-------------------------------------
Because of the analytical description that we achieved, \textit{Chirikov's diffusion}, the diffusion of an orbit along a resonant chain (a consequence of the overlapping criterion \cite{aMo95}), was natural to suspect.
In order to measure the value of the diffusion coefficient, we introduce the  mean-squared displacement in eccentricity, 
\begin{align}
\sigma^{2}(\tau)~\equiv~\big\langle \big(\Delta e(\tau) - \langle \Delta e(\tau) \rangle\big)^{2}\big\rangle; 
\end{align}
the diffusion coefficient related to the eccentricity being defined as
\begin{align}\label{eq:De}
	\mathcal{D}_{e}(\tau) \equiv \lim_{\tau \to +\infty} \frac{\big\langle \big(\Delta e - \langle \Delta e \rangle\big)^{2}\big\rangle}{2\tau},
\end{align}
where $\Delta e(\tau) = e_{t_{i}+\tau} - e_{t_{i}}$ and $\langle \bullet \rangle$ is an average operator.
We have computed the coefficient $\mathcal{D}_{e}(\tau)$ in a purely numerical way following  orbits for long timespans\footnote{More than $5$ centuries, this timescale represents around $3.5 \times10^{5}$ revolutions around the Earth.} using our precision orbit propagator. This type of averaging is called \textit{dynamic averaging}\footnote{
The \textit{dynamic averaging} differs from the \textit{spatial averaging}, where the average operator is over some appropriate 
ensemble, but they gave the same results if the ergodic hypothesis holds.  
}$^{,}$\cite{aRe83}, and the coefficient $\mathcal{D}_{e}$ is quantitatively defined as 
\begin{align}
	\mathcal{D}_{e}(\tau) = \lim_{n \to \infty} \lim_{\tau \to \infty} \frac{1}{2 \tau} \frac{1}{n} \sum_{i=1}^{n} \big( (e_{(i+1)\tau} -e_{i\tau})-\overline{\Delta e} \big)^{2}.
\end{align}

\begin{figure}
  \centering
\includegraphics[scale=0.4]{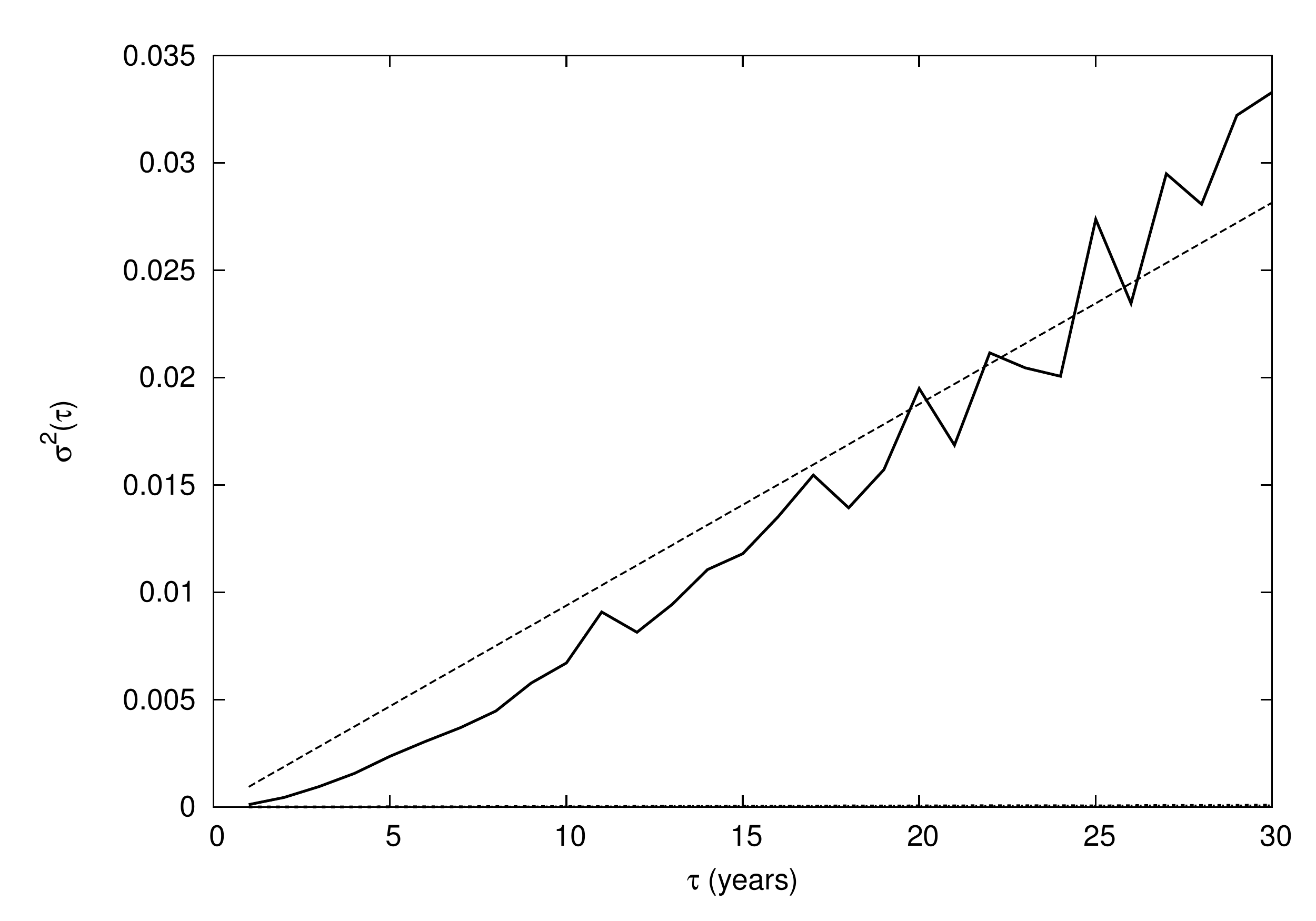}
\caption{\label{fig:D-versus-time}
Time evolution of $\sigma^{2}(\tau)$ fitted by linear least-squares curves for the $2$ orbits of Fig.\,\ref{fig:e-versus-time}.
For the  regular orbit the curves are flat and indistinguishable from the $x$-axis, while for the chaotic case, a linear trend with a non-zero slope is well captured.  
}
\end{figure}

Figure\,\ref{fig:D-versus-time} shows the evolution of the mean-squared displacement as a function of the length $\tau$ for the $2$ orbits of Fig.\,\ref{fig:e-versus-time}. 
Firstly, it is legitimate to talk about \textit{diffusion}: diffusive processes are commonly characterized by a power law relationship 
\begin{align}
	\sigma^{2}(\tau) \propto \tau^{\nu}.
\end{align}
Having here that $\nu$ is very close to $1$, we found a \textit{normal} diffusion behavior for the chaotic orbit.
We found also that the diffusion coefficient changes significantly depending on the orbit's nature;  
the slopes of the linear least-squares fit  differ up to $6$ orders of magnitude for the  chaotic and the regular orbit, which appears here as flat. 
%----------------------------------
%---------------------------------

\begin{figure}
\centering
\begin{tabular}{cc}
\vspace{-0.3cm}
\hspace{-1.1cm}
\includegraphics[scale=0.25]{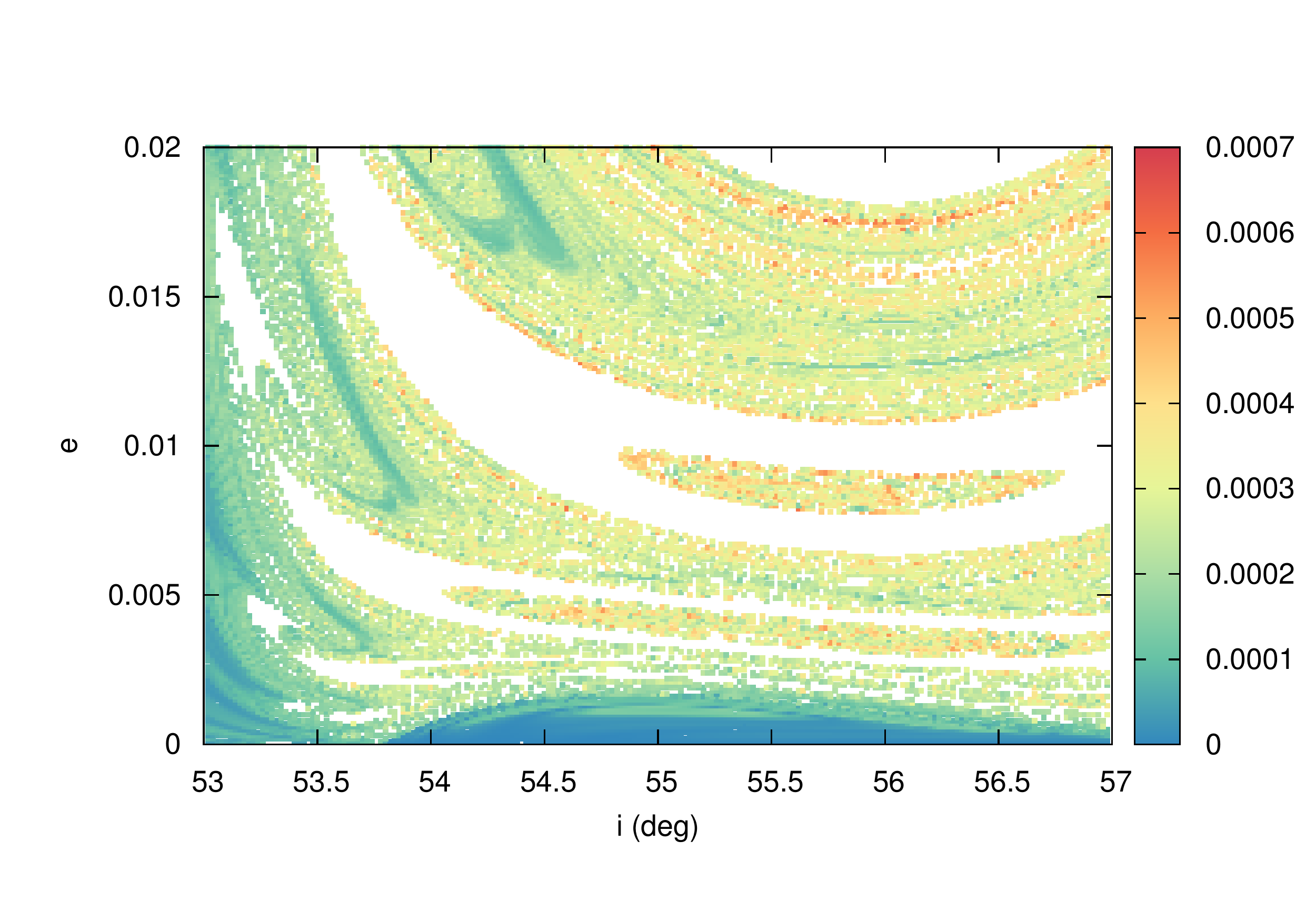} &
\hspace{-0.6cm}
\includegraphics[scale=0.25]{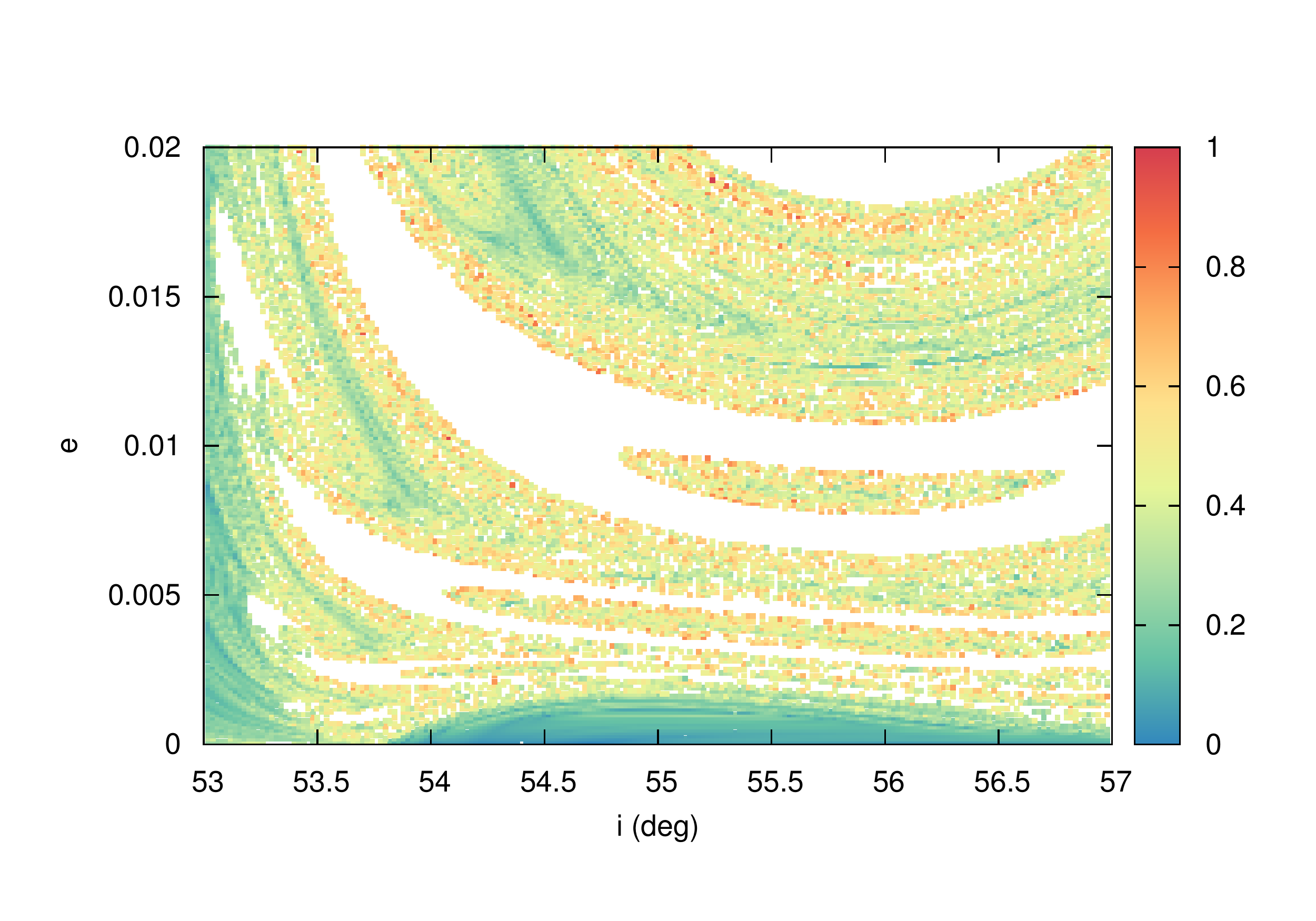} \\
\hspace{-1.1cm}
{\small(a) Diffusion map.} & 
\hspace{-0.6cm}
{\small(b) FLI map.}
\end{tabular}
\caption{\label{D-map1} 
Stability maps characterizing the diffusivity and the local hyperbolicity (normalized to $1$) for physical parameters relevant to the MEO region computed as a function of the initial eccentricity and inclination. Initial phases of the system are $\Omega=120^{\circ}$ and  $\omega=120^{\circ}$.}
\end{figure}

\begin{figure}
\centering
\includegraphics[scale=0.4]{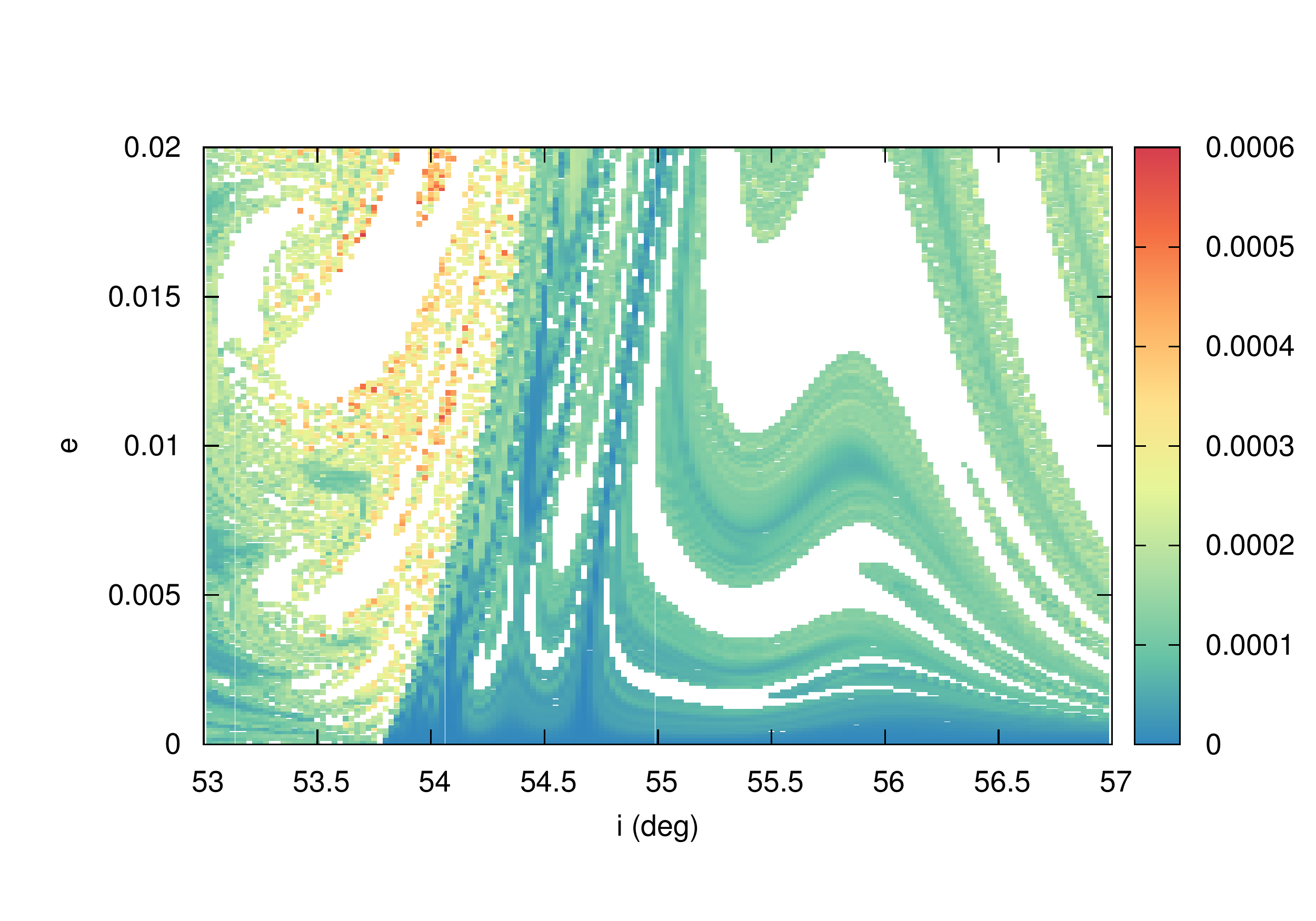}
\vspace{-0.3cm}
\caption{\label{D-map2} Diffusion map for physical parameters relevant to the MEO region computed as a function of the initial eccentricity and inclination. Initial phases of the system are $\Omega=240^{\circ}$ and  $\omega=120^{\circ}$.}
\end{figure}

We extended the computation of $\mathcal{D}_{e}$ for a particular domain of the $i$--$e$ phase-space. Figures \,\ref{D-map1} shows the results of the computation in the MEO region, for physical parameters relevant to navigation satellites, covering a small domain of the phase-space; namely, 
the rectangle $[0,0.02] \times [53^{\circ},57^{\circ}]$, spaced uniformly by $185 \times 160$ initial conditions. The palette scale gives the magnitude of $\mathcal{D}_{e}$, indicating in which phase-space region the diffusivity is fastest. For the volume of the phase-space that we have explored up to a large, but finite time $t_{\rm f}$, all  diffusion coefficients were finite, an indication  that the motion does not spread more rapidly than diffusively.  
This lead to the important fact that typical navigation satellites obey reasonably well to a diffusion law. When KAM curves are approached, we found that the diffusion coefficients goes to zero sufficiently fast.  Namely, by comparing the results of the diffusion maps with the FLI maps computed for the same physical parameters, we found a very nice agreement between the dynamical structures revealed either by $\mathcal{D}_{e}$ or the FLIs, implying in general a  $1$--$1$ correspondence between high local hyperbolicity and high diffusivity\footnote{A non-trivial result due to the existence of \textit{stable chaos}.} as shown in Fig.\,\ref{D-map1}. 
It is important to note that even for moderate eccentricity, $e \le 0.005$, we may find high diffusivity regions, whose spatial organization in the phase-space is complex. 
We redo the same computation as that in Fig.\,\ref{D-map1}, but we change the initial phases of the system (all others parameters are identical), as presented in Fig.\,\ref{D-map2}. We can observe how the structures evolve by changing the initial angles, even if in both cases high diffusive orbits can be found.  This observation illustrates in essence the difficult question of determining which initial phases of the initial state  leads to a diffusive chaotic response of the system. This is intimately related to the representation of the dynamics in a reduced dimensional phase-space.  Moreover, the diffusion coefficient calculated here give no information about which angles will ensure or avoid diffusive chaos for a fixed initial eccentricity and inclination. This point, of particular practical interest,  undoubtedly needs further investigation. At the very least, diffusion maps as those presented here should be computed for an ensemble of initial phases\footnote{Note that the ergodic nature of the phase-space has not yet been investigated.}, but this point represent a difficult and formidable computational task. Let it be recalled that the Hamiltonian of Eq.\,\eqref{Hbm} is $3$-DOF and autonomous, the global understanding of such systems being on the cusp of current trends in dynamical systems research. 

%----------------------------------
%----------------------------------

%----------------------------------
%----------------------------------

%-------------------------------------
\section{Conclusions}
%-------------------------------------
The overlapping of lunisolar secular resonances in MEO gives rise to complex chaotic dynamics affecting mainly the eccentricity. The local hyperbolicity associated to the resulting stochastic layer is synonymous to macroscopic transport in action space, with typical Lyapunov times on the order of decades. We have shown that these transport properties obey a diffusion law, and we have presented dynamical maps based on the numerical estimation of the diffusion coefficient.
Our results show that we may find diffusive orbits even for moderate eccentricity near the operational inclinations of (and for physical parameters relevant to) the navigation satellites. Nonetheless, the number of degrees of freedom renders difficult the global comprehension of the tableau, as attested by the diffusion maps presented herein. The computational challenge emanating from this difficulty is surely a nice invitation to go beyond our communities, and to make considerable efforts to redesign the standard way by which `stability maps' in general are traditionally generated in Celestial Mechanics. 
Significative improvements will arise certainly using adaptive algorithms and non-structured grids, based on clustering techniques \cite{nNa15}. 
 
% Unnumbered appendix sections can be obtained using \verb|\section*|.

\bibliographystyle{ws-procs9x6} % for numbered citation & references
%\bibliography{ws-pro-sample}

%\end{document}

\end{document}